\documentclass{elsart}
\usepackage{graphicx}
\usepackage{amssymb}

\begin{document}

\begin{frontmatter}


\title{Neutron beam test of CsI crystal for dark matter search}

\author[SNU,DMRC]{H. Park\corauthref{cor1}}
\corauth[cor1]{Corresponding author: Tel 82-2-876-2801  Fax 82-2-875-4719 e-mail hyeonseo@hep1.snu.ac.kr}
\author[SKK]{D. H. Choi}
\author[SNU]{J. M. Choi}
\author[Ewha]{I. S. Hahn}
\author[Yonsei]{M. J. Hwang}
\author[Sejong]{W. G. Kang}
\author[Yonsei,DMRC]{H. J. Kim}
\author[Ewha]{J. H. Kim\corauthref{JHKIM}}
\corauth[JHKIM]{present address : Department of Physics, University of Texas at Austin, Austin, Tx 78712, USA}
\author[SNU]{S. C. Kim}
\author[SNU]{S. K. Kim}
\author[SNU,DMRC]{T. Y. Kim}
\author[Sejong]{Y. D. Kim}
\author[Yonsei]{Y. J. Kwon}
\author[SNU]{H. S. Lee}
\author[Yonsei]{J. H. Lee}
\author[UMIPST]{M. H. Lee}
\author[SNU]{S. E. Lee}
\author[SKK]{S. H. Noh}
\author[SNU]{I. H. Park}
\author[UM]{E. S. Seo}
\author[SNU]{E. Won}
\author[SNU]{H. Y. Yang}
\author[SNU]{M. S. Yang}
\author[SKK]{I. Yu}
\address[SNU]{Department of Physics, Seoul National University, Seoul
151-742, Korea} 
\address[DMRC]{Dark Matter Research Center, Seoul National University, Seoul
151-742, Korea} 
\address[SKK]{Physics Department, Sungkyunkwan University, Suwon
440-746, Korea}
\address[Ewha]{Department of Science Education, Ewha Woman's
University, Seoul 120-750, Korea}
\address[Yonsei]{Physics Department, Yonsei University, Seoul 120-749, Korea}
\address[Sejong]{Department of Physics, Sejong University, Seoul
143-747, Korea} 
\address[UMIPST]{Institute of Physical Science and Technology, University of Maryland, College Park, MD 20742, USA}
\address[UM]{Physics Department, University of Maryland, College Park,
MD 20742, USA}

\begin{abstract}
We have studied the response of Tl-doped and Na-doped CsI crystals to
nuclear recoils and $\gamma$'s below 10 keV. The response of CsI
crystals to nuclear recoil was studied with mono-energetic neutrons
produced by the $^3$H(p,n)$^3$He reaction. This was compared to the
response to Compton electrons scattered by 662 keV $\gamma$-ray.
Pulse shape discrimination between the response to these $\gamma$'s
and nuclear recoils was studied, and quality factors were
estimated. The quenching factors for nuclear recoils were derived for
both CsI(Na) and CsI(Tl) crystals.
\end{abstract}

\end{frontmatter}

\newpage
% main text
\section{Introduction}

It is known that a major component of matter in the Universe is not
ordinary matter(luminous stars and/or baryons) but nonbaryonic exotic
matter\cite{ref:DarkMatter}. The Weakly Interacting Massive
Particle(WIMP) is one of the strongest candidates for this dark
matter, and direct searches for WIMPs using various detection
techniques have been undertaken
\cite{ref:DAMA,ref:UKDMC,ref:CDMS,ref:CRESST,ref:EDELWISS,ref:HEIDEL}.

The sensitivity of WIMP detection strongly depends on various
parameters of the detector, such as detection threshold, mass of the
detector, background rate, pulse shape discrimination(PSD)
capabilities, {\it etc}.  A WIMP may interact with the target nuclei
through a WIMP-nucleus elastic scattering, and the detection of the
nuclear recoil signal may give a signature of WIMP detection. The
expected recoil energy is several tens of keV, depending on the masses
of the target nucleus and the WIMP. However, because of the high
stopping power of the recoil nucleus, the light yield of a
scintillation detector is 5-10 times lower in comparison to a signal
from a $\gamma$ of equivalent energy due to quenching effect in normal
crystals. Thus, the detection threshold should
be low enough to measure a few keV of energy.

The expected event rate of the WIMP-nucleus interaction is less than 1
event /kg/day. Such a low event rate requires an extremely low
background rate, on the order of several counts/kg/day/keV, and a
large detector size in order to increase the signal rate. To reduce
the background rate, the PSD between nuclear recoils and background
$\gamma$'s or electrons in the several keV region, can be utilized.

A CsI crystal is a good candidate for a WIMP detector thanks to its
large light yield and relatively good PSD capability. Also, it is
relatively easy to fabricate large detector volumes because the
crystals are not hygroscopic. The large masses of Cs and I nuclei
enhance the cross-section for a WIMP-nucleus, spin-independent
interaction(which depends on  $A^2$ \cite{ref:A2}).  However, the
detailed response of the crystal to low energy radiation has not been
fully investigated. Recently, the feasibility of using CsI(Tl)
crystals as a WIMP detector was shown\cite{ref:KIMS}, while several
other groups reported the characteristics of a CsI(Tl) crystal for
WIMP detection\cite{ref:Pecourt,ref:OTHERS}.

We studied here the response of CsI crystals with several different
doping concentrations of Tl and Na to neutrons and $\gamma$'s. The
energy region was several keV for the $\gamma$'s and several tens keV
for the nuclear recoils. In this paper, the measurements of PSD and the
quenching effect of the nuclear recoils for both CsI(Tl) and CsI(Na)
crystals are reported.

\section{Experiments}

The response of CsI crystals to neutrons was studied using a
tandem accelerator(maximum terminal voltage of 1.7 MV) at the Korea
Institute of Geoscience and Mineral resources(KIGAM). 
Mono-energetic neutrons of 2.62 MeV were produced at zero degrees via
the $^3$H(p,n)$^3$He reaction, using a 7 to 9 nA beam of 3.4 MeV
protons on a $^3$H target. These neutrons produce nuclear recoils
with energies near that expected for WIMP interactions.

\begin{figure}[htb]
\begin{center}
\includegraphics[height=12cm]{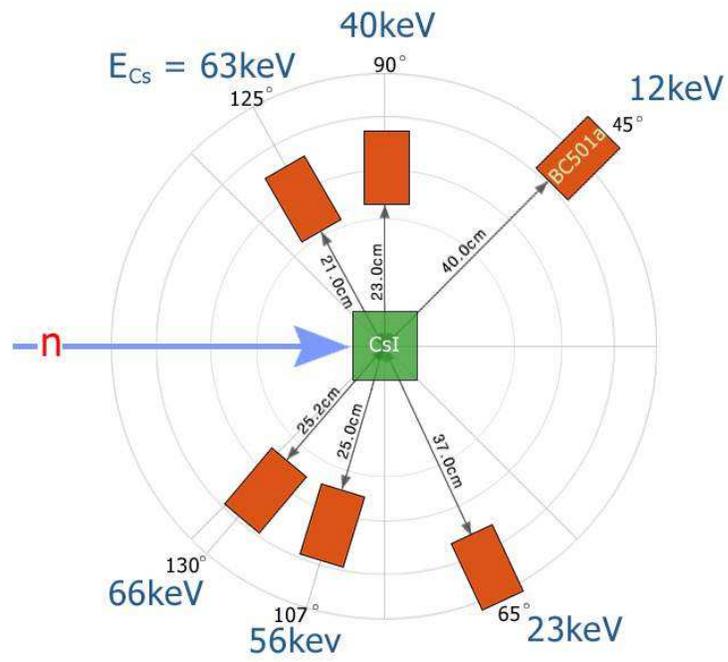}
\end{center}
\caption{Experimental Setup.}
\label{fig:setup}
\end{figure}

Figure \ref{fig:setup} shows a schematic of the experimental
setup. A CsI crystal was located at zero degrees with respect to the
proton beam line. The CsI crystal was 3 cm $\times$ 3 cm $\times$ 3 cm, in
cubic shape, and with all 6 faces polished. Two 3-inch PMTs(D726UK,
Electron Tubes Limited), with green-extended RbCs bi-alkali
photocathodes, were directly attached to the top and bottom surfaces.
The top and bottom sides of the crystal were completely
covered by the PMT cathode plane, and the other four sides were wrapped
with two layers of Teflon (0.2 mm thick) followed by black vinyl
sheets. In order to identify neutrons scattered from CsI, six
neutron detectors were located at various angles ranging from 45 to
130 degrees. The detailed locations of the neutron detectors are shown
in Figure \ref{fig:setup}. The neutron detectors were made of liquid
scintillator(BC501A) contained in a cylindrical Teflon container of
12.7 cm diameter $\times$ 10 cm long. One 2-inch PMT(H1161, Hamamatsu
Photonics) was used for readout. Each neutron detector was contained
in a polyethylene block, about 5 cm thick, and was shielded with lead
and borated paraffin blocks.  The recoil energy($E_{recoil}$) of the
nucleus can be calculated by a simple kinematical equation using the
incident neutron energy and the scattering angle of the neutron, as 
determined from the position of neutron detector:
\begin{equation}
E_{recoil} = E_{beam}\cdot\left\{1-\left({m_ncos\theta - \sqrt{m_N^2 - m_n^2\sin^2\theta}\over m_n + m_N}\right)^2\right\} ,
\label{eq:Erecoil}
\end{equation}
where $E_{beam}$ is neutron energy, $m_n$ and $m_N$ are the masses of
neutron and recoiling nucleus(Cs or I)\footnote{In the real
calculation, the average
mass of Cs and I was used.} respectively, and $\theta$ is the neutron
scattering angle.

The signals from a CsI crystal are amplified using home-made, fast
amplifiers ($\times$8) with low noise and high slew rates. One output
from these amplifiers ($\times$8) was directly connected to a digital
oscilloscope(LeCroy LT364) in order to record the pulse shape. The other signal
is amplified ($\times$10), again using a Philips amplifier(Philips 770),
and connected to a discriminator(Philips 711) to form the trigger
logic. The discriminator threshold was set to the level of single
photoelectrons. In order to suppress the accidental trigger from
PMT dark current, at least 4 photoelectrons within 1 $\mu s$(actually
2 photoelectrons within 0.5 $\mu s$ for each PMT) were
required(a so-called ``4-fold'' coincidence). The detailed description
is shown in the reference\cite{ref:KIMS}.

The event rate of 4-fold coincidences of the signals from the crystal 
was about 1200 Hz when the neutron beam was on. Most of these events were due
to the $\gamma$ backgrounds induced by the neutron beam.  In order to
reduce the trigger rate, a time coincidence of a CsI signal(4-fold)
and the neutron detector signal must occur within 500 ns. This reduced
the trigger rate to less than 3 Hz. Trigger timing was
determined by the neutron detector. The CsI signal was digitized by the
digital oscilloscope(LeCroy LT364) with a 100 MHz sampling frequency and
a 10 $\mu s$ full range. The trigger timing for the oscilloscope was set at
3 $\mu s$. The neutron detector signal was digitized by a charge
sensitive camac
ADC(LeCroy 2249A). Data acquisition used ROOT\cite{ref:ROOT} on a
linux machine. 

In order to compare the pulse shape of a nuclear recoil to that of a
$\gamma$, the crystal was irradiated with a $^{137}$Cs source to
induce Compton scattering. This induced a signal which is distributed
uniformly inside the crystal. These data were taken separately using
the same setup, but without the proton beam. The trigger was a 4-fold
coincidence from a crystal.

Table \ref{tab:crystal} shows a list of the crystals tested. Two
doping materials, Na and Tl, with various concentrations, were
tested. The crystals were supplied by the Institute for Single
Crystals in Kharkov, Ukraine. Growing methods of each crystal are
indicated in the table.

The doping concentrations of the CsI(Tl) crystals were measured by
the ICP-AES (Inductively Coupled Plasma - Atomic Emission Spectrometry)
method. The 4 corners of each crystal were scratched by a razor blade,
and sent to two independent institutes. Each institute measured two
samples for each crystal. The errors of the doping concentration shown
in the table are the standard deviation of the 4 measurements.

For CsI(Na), the doping concentrations were measured by
ICP-MASS (Inductively Coupled Plasma - Mass Spectrometry) method.  The
measurement was done by only one institute and the measurement errors
were claimed to be 2-3 \%.

\section{Analysis}

\subsection{Typical low energy signal from a CsI crystal}

The single photoelectron(SPE) signal from a PMT has less than a 10 ns
full width. After amplification, the signal is stretched a bit due to
the inherent shaping time of the amplifier(the rise time is about 10 ns
and the fall time is about 20 ns). Hence, the SPE signal spreads
about 30-40 ns in the time spectrum.  It is known that the decay time
of CsI(Tl) is about 1 $\mu$s \cite{ref:Knoll}. Since the number of
photoelectrons for low energy events (less than 10 keV where WIMP
signals are expected to occur) was less than 60 (or 30 per one PMT),
the SPEs were
clearly separated from each other in an event.  That is, a low energy
real event can be identified as a set of SPEs above a 
threshold cut within a timing window of 10 $\mu$s.

\begin{figure}[htb]
\begin{center}
\includegraphics[width=10cm]{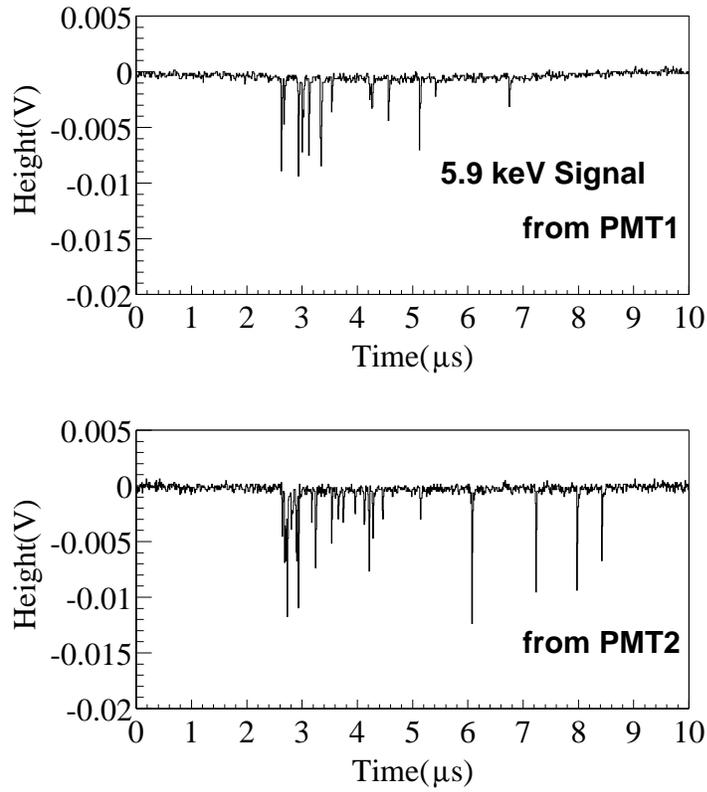}
\end{center}
\caption{Typical signal from CsI(Tl) crystal for two PMTs
obtained by 5.9 keV X-ray from $^{55}$Fe source.}
\label{fig:5.9keV}
\end{figure}

An example of a low energy signal is shown in Figure \ref{fig:5.9keV}.
This signal was produced by 5.9 keV X-ray from a $^{55}$Fe source.  The
average number of photoelectrons from a CsI(Tl) crystal was about 5
p.e./keV. As expected, the signal was composed of a set of well
separated peaks composing the SPE signal. In order to reduce
the noise contribution, a 'clustering algorithm' was developed to
identify SPEs. The pedestal of each event was estimated in the first 2
$\mu s$ window before the trigger for each PMT. Each peak of about a
30-40 ns width and above a threshold value was considered one cluster.

\begin{figure}[htb]
\begin{center}
\includegraphics[width=12cm]{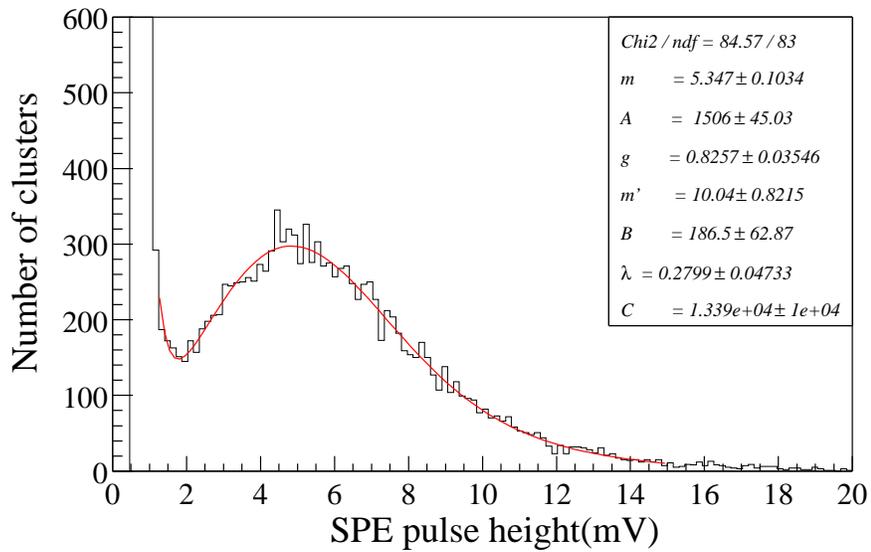}
\end{center}
\caption{Single photoelectron spectrum: The fit results are
overlaid(solid curve), done using two Poisson functions.}
\label{fig:SPE}
\end{figure}

Figure \ref{fig:SPE} shows the pulse height distribution of the
clusters composed of having 4 keV $< E_{meas} <$ 8 keV, where
$E_{meas}$ is the measured electron-equivalent energy. The figure
shows separation between the SPEs and the noise,
with a peak to valley ratio of about 2. If the cluster
is a real SPE, the pulse height distribution should follow a Poisson
distribution, because the 
amplification at the first dynode inside the PMT follows Poisson
statistics. In an attempt to understand the shape of the spectra, a
fit was done using two superimposed Poisson
functions(one for the SPE and the other for the SPE-overlapped signal) and
an exponential function for noise, as expressed in the following equation.
\begin{equation}
f = A \cdot {\mu^r \cdot e^{-\mu} \over \Gamma(r+1)} +  B \cdot
{{\mu^\prime}^r \cdot e^{-\mu^\prime} \over \Gamma(r+1)} +C \cdot
e^{-x/\lambda},
\label{eq:poisson}
\end{equation}
\[ r = x\cdot g, \mu = m\cdot g, \mu^\prime = m^\prime \cdot g,\]
where $x$ is the pulse height, $m(m^\prime)$ is a number
corresponding to the mean of the Poisson distribution, and $g$ is the gain
factor of the PMT(from the second dynode). The results of the fit are
overlaid in the figure as a solid line, where the reduced $\chi^2$ of
the fit is about 1. The ratio of the mean values of two Poisson
distributions, $m/m^\prime$, was 1.88, compared to an expected value of 2. The
ratio of the contributions of the two Poisson distributions is about
12\%. So, we could conclude that the cluster is a SPE ($\sim$ 80\%) or
due to overlapped photoelectrons($\sim$ 20\%).
The same fit was performed for events of 2 keV $< E_{meas} <$ 4 keV.
A single Poisson function will fit the SPE spectrum.

\subsection{Neutron identification}

\begin{figure}[htb]
\begin{center}
\includegraphics[width=10cm]{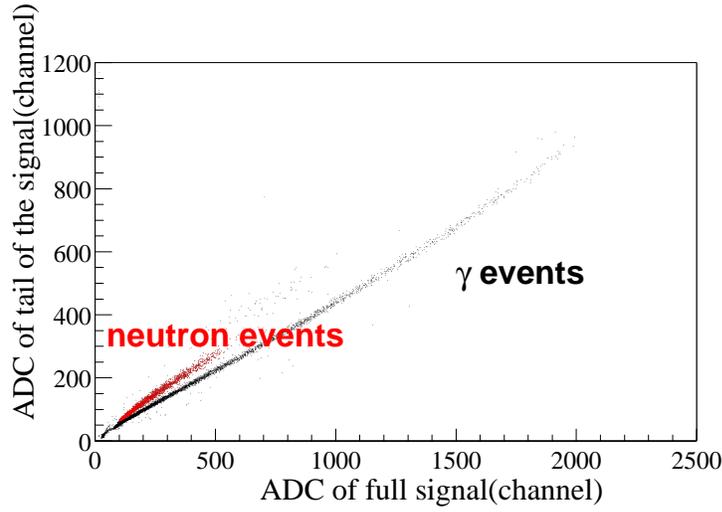}
\end{center}
\caption{Neutron identification plot for neutron detector.}
\label{fig:nid}
\end{figure}

In order to identify neutron induced events, we required a
coincidence between one of the 6 neutron detectors and the CsI crystal.
Furthermore, neutron scattering events can be
selected by separating neutrons and $\gamma$'s in the neutron
detector.  It is well known that liquid scintillator(BC501A) can
separate neutrons from $\gamma$'s using PSD. The neutron(proton recoil)
events have a longer tail than $\gamma$ events for the same energy
loss\cite{ref:Knoll} in the BC501A liquid scintillator. Therefore, using
the ratio of the total charge to the charge in the tail of the
signal, the neutrons can be clearly separated from $\gamma$'s. Figure
\ref{fig:nid} shows the neutron identification plot for the neutron
detector. Above 300 keV, the neutron-$\gamma$ separation is about
2.7$\sigma$ which is sufficient for neutron-event
selection. The detailed description for neutron detector will be
described in a separate report\cite{ref:NeutronDetector}.

\begin{figure}[htb]
\begin{center}
\includegraphics[width=15cm]{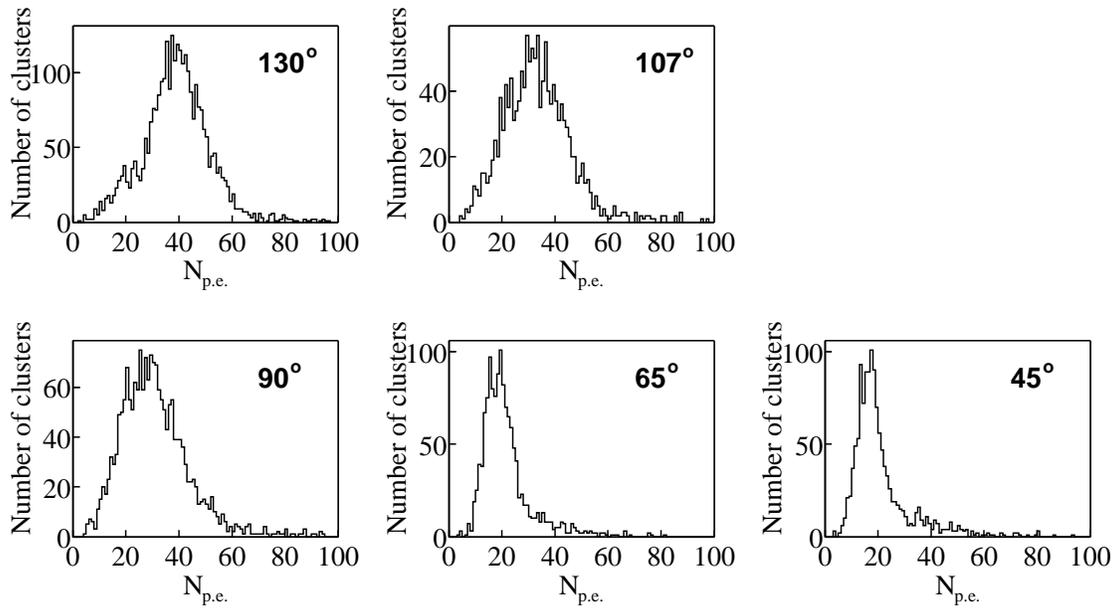}
\end{center}
\caption{Recoil energy spectra of CsI(Tl-0.073mole\%) tagged by
each neutron detector. ${\rm N_{p.e.}}$ in the x-axis indicates the number
of photoelectrons.}
\label{fig:Erecoil}
\end{figure}

Figure \ref{fig:Erecoil} shows the recoil energy spectra of the
CsI(Tl-0.073mole\%) crystal tagged by a neutron detector. The measured
energy increases as the scattering angle increases as expected. A
quantitative analysis will be described in Section \ref{sec:QuenchingFactor}.

\subsection{Timing characteristics of CsI crystal}

\begin{figure}[htb]
\begin{center}
\includegraphics[width=10cm]{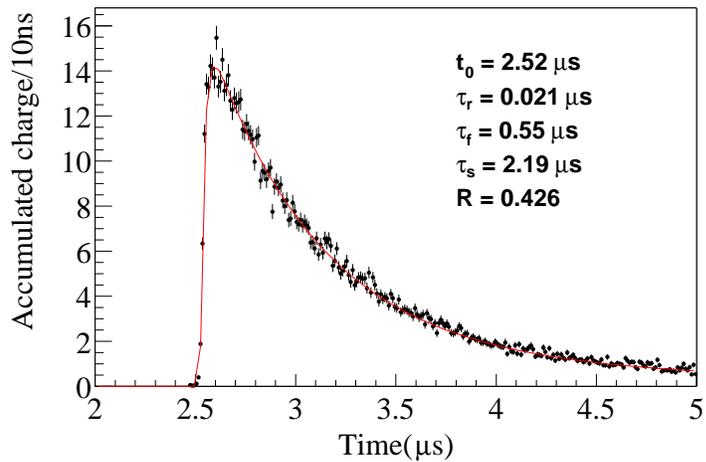}
\end{center}
\caption{Time spectrum from CsI(Tl-0.073mole\%). The relative
time of each event is fixed by one neutron detector at 3 $\mu$s. The
solid curve represents the fit result by Eq. \ref{eq:risingtime}.}
\label{fig:risingtime}
\end{figure}

To study the overall timing characteristics of a crystal at low
energy, we accumulated whole events for each crystal.  Figure
\ref{fig:risingtime} shows the accumulated time spectrum of a
CsI(Tl-0.073mole\%) crystal for neutron induced events, whose measured
energy is 4 keV $< E_{meas} <$ 10 keV. In the figure, the time scale
is in $\mu s$,  and the timing of the neutron detector was set to 3
$\mu s$.

The timing characteristics of the CsI response can be parameterized by
a rise time($\tau_r$), two decay times($\tau_f$ and $\tau_s$), and the
ratio of slow to fast decay components($R$): 
\begin{equation}
F(t) = A\cdot \left\{{1\over \tau_f} \cdot e^{-{t-t_0\over \tau_f}} +
{R\over \tau_s} \cdot e^{-{t-t_0\over \tau_s}} - \left({1\over \tau_f} +
{R\over \tau_s}\right)\cdot e^{-{t-t_0\over \tau_r}}\right\} + b.g.,
\label{eq:risingtime}
\end{equation} 
where time zero($t_0$), overall normalization($A$), and the amount of
background photoelectrons($b.g.$) are included. The amount of
background photoelectrons was estimated from the number of
photoelectrons within a 2 $\mu$s window during the 1 $\mu$s prior
to the event. The constant $b.g.$ values were obtained by averaging all
the events under analysis, resulting in about 0.1-0.3 photoelectrons
per 1 $\mu$s. The fit was over 10 $\mu$s even though the figure shows
only 5 $\mu$s. The values of parameters from the fit are shown in
Table \ref{tab:timing}. For the Tl-doped crystals, the rise time is
about 20 ns and it has two components of decay times, fast(about 0.54
$\mu$s) and slow(about 2 $\mu$s) components. For the Na-doped
crystals, the rise time is about 30 ns, the fast decay component is
about 0.41 $\mu$s, and the slow decay component about 2.4 $\mu$s. The time
characteristics of CsI(Na) are similar to those of CsI(Tl) at low
energy. This is not an expected feature because it is well known that the
decay time of CsI(Na) is much shorter than that of CsI(Tl) at high
energy\cite{ref:Knoll}.

We do not observe a significant
dependence on the doping concentration
except for the case of CsI(Tl-0.071mole\%). Both
CsI(Tl-0.071mole\%) and CsI(Tl-0.073mole\%) have the same doping
material and the same concentration, and the only known difference is
the growing method, Kyropoulos(CsI(Tl-0.071mole\%)) and
Bridgman(CsI(Tl-0.073mole\%)). However, it is not clear if 
different doping methods cause different time characteristics because
we tested only one CsI(Tl) crystal made by Kyropoulos method.

\section{Pulse Shape Discrimination}

\subsection{Comparison between nuclear recoil and $\gamma$ signals}

\begin{figure}[htb]
\begin{center}
\includegraphics[width=10cm]{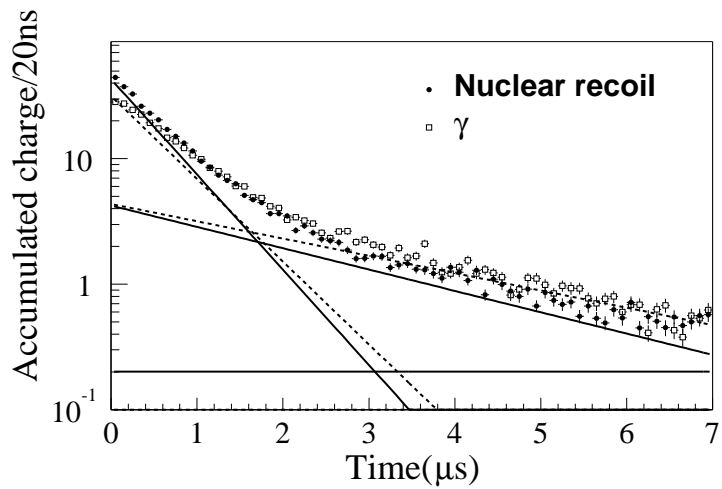}
\end{center}
\caption{Time spectra from CsI(Tl-0.073mole\%) for nuclear recoil and
$\gamma$ events. The timing of the first photoelectron of each event
is fixed at 0. Each decay component and the background are decomposed
by a fit to Eq. \ref{eq:decayspec}. The solid lines are for
nuclear recoil data and the dashed lines are for $\gamma$ data, where
the background for $\gamma$ events is less than 0.1. For a better
display, only one out of every 5 points in the X-axis is shown in the
figure.}
\label{fig:decayspec}
\end{figure}

Pulse shape discrimination(PSD) with a CsI crystal is possible due
to different timing characteristics between neutron induced (nuclear
recoil) and $\gamma$ induced events\cite{ref:PSD}.  Figure \ref{fig:decayspec}
shows the accumulated time spectrum of a CsI(Tl-0.073mole\%) crystal for
4 keV $< E_{meas} <$ 10 keV, where the time spectrum for $\gamma$
events and that for nuclear recoil events are overlaid. For each
event, the timing of the first photoelectron was used as $t_0$.
The ambiguity due to improper determination of $t_0$
should affect the time distribution, especially for the low energy
event which does not have a sufficient number of photoelectrons.
Eq. \ref{eq:decayspec} shows the fit function for the time spectrum,
which is similar with Eq. \ref{eq:risingtime}. In order to reduce the
bias on the $t_0$ determination, the fit was done over 0.3
$\mu$s. Each component of the time spectrum is drawn as a solid
line(nuclear recoil) or a dashed line($\gamma$). The amount of
background photoelectrons for both cases are also shown. The fit shows
that nuclear recoil events clearly have a shorter tail than $\gamma$
events.

\begin{equation}
F(t) = A\cdot \left({1\over \tau_f} \cdot e^{-{t\over \tau_f}} + {R\over \tau_s} \cdot e^{-{t\over \tau_s}}\right) + b.g.
\label{eq:decayspec}
\end{equation}

\subsection{Quality factor}

To quantify the capability for neutron-$\gamma$ discrimination among
the different crystals, and to evaluate the sensitivity of a WIMP search,
a quality factor was defined\cite{ref:QualityFactor} as
\begin{equation}
K \equiv {\beta\cdot (1-\beta) \over (\alpha - \beta)^2},
\label{eq:QFdefine}
\end{equation}
where $\alpha$ is the fraction of signal events passing the event
selection criteria and $\beta$ is the fraction of background events
which 
passed the same criteria. For an ideal detector, $\alpha = 1$ and
$\beta = 0$. Therefore, a smaller quality factor means a better
separation between signal and background events.

\begin{figure}[htb]
\begin{center}
\includegraphics[height=8cm]{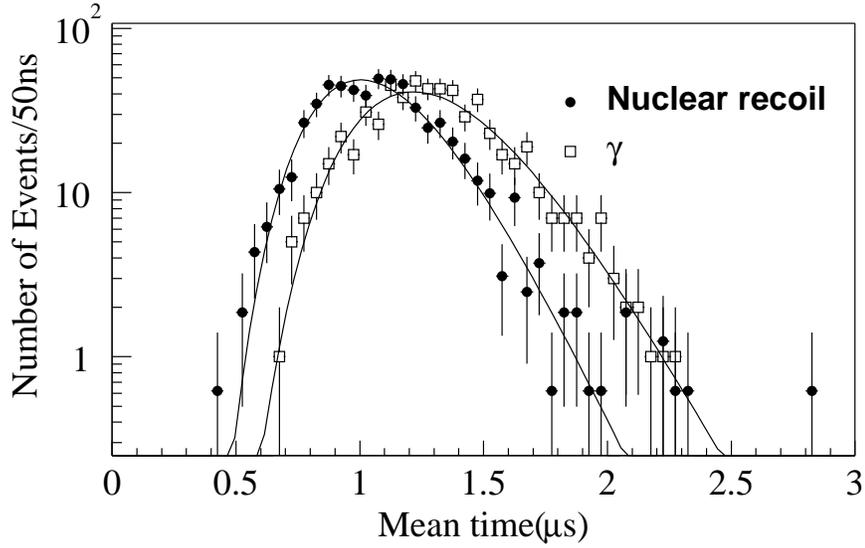}
\end{center}
\caption{Mean time distribution for CsI(Tl-0.073mole\%) for
nuclear recoil and $\gamma$ events. The mean time distributions are
fitted well with the Log-Gaussian function(solid curves).}
\label{fig:MeanTime}
\end{figure}

For a simple separation parameter, a mean time is defined as;
\begin{equation}
<t> = {\sum A_i t_i \over \sum A_i} - t_0,
\end{equation}
where $A_i$ is the charge of the $i$-th cluster, $t_i$ is the time of
the $i$-th cluster, and $t_0$ is the time of the first cluster(assumed
as time zero). Figure \ref{fig:MeanTime} shows the mean time
distribution for nuclear recoil and $\gamma$ events for a
CsI(Tl-0.073mole\%) with events between 6 keV $ <E_{meas}<$ 7 keV. 
Nuclear recoil and $\gamma$ events show different mean
times. By selecting the proper signal window, we can choose the event
sample which maximizes the signal to background ratio.

\begin{figure}[htb]
\begin{center}
\includegraphics[height=8cm]{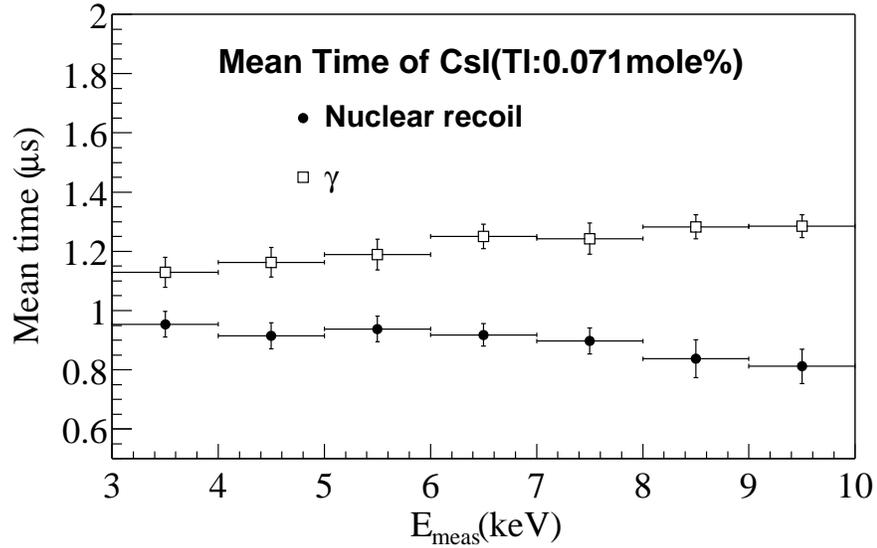}
\end{center}
\caption{Mean time of CsI(Tl-0.071mole\%) for each energy
bin, calculated from the time constant of each crystal. The errors in
Y-axis are statistical only. The error bars in X-axis show the energy
region used for each data point.}
\label{fig:comparemt}
\end{figure}

However, due to background photoelectrons which are uniformly
distributed, the mean value of the $<t>$ distribution is longer than
the real mean time. To estimate the effect of background
photoelectrons, the mean value of the mean-time distribution is
calculated with the time constants($\tau_f$, $\tau_s$, and $R$)
obtained from the fit shown in Eq. \ref{eq:decayspec} for each energy
bin. The variance of the mean time distribution is calculated by removing
the contribution from the background photoelectrons by assuming a
quadratic sum($\sigma_{<t>}^2 = \sigma_{signal}^2 + \sigma_{b.g.}^2$),
where $\sigma_{<t>}^2$, $\sigma_{b.g.}^2$, and $\sigma_{signal}^2$ are
the variance of the $<t>$ distribution, the background photoelectrons,
and the signals, respectively. Figure \ref{fig:comparemt} shows the
mean time of the nuclear recoil and $\gamma$ events calculated from the
time constant of the crystal. The errors are only statistical.

\begin{figure}[htb]
\begin{center}
\includegraphics[width=10cm]{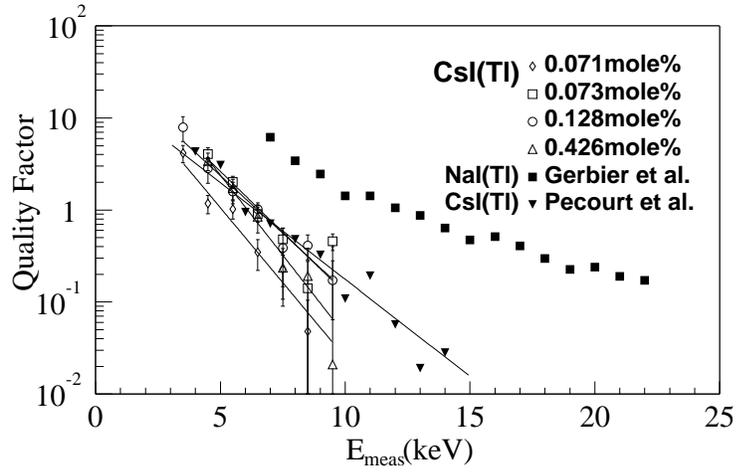}
\end{center}
\caption{Quality factors for various CsI(Tl). The errors are only
statistical. The present results(open markers) are compared to the
data of S. P\'{e}court {\it et al.}\cite{ref:Pecourt} and of
G. Gerbier {\it et al.}\cite{ref:PreMeasKFactor}.}
\label{fig:quality}
\end{figure}

Figure \ref{fig:quality} shows the quality factors for CsI(Tl) for
various doping concentrations. The results from previous
measurements for CsI(Tl)\cite{ref:Pecourt} and for
NaI(Tl)\cite{ref:PreMeasKFactor} are overlaid.  For CsI(Tl), the
present results show a bit more energy dependence than
Ref. \cite{ref:Pecourt}. However, in the energy region
less than 5 keV where most of the WIMP signals are expected, both
measurements are comparable. CsI(Tl) shows in comparison to NaI(Tl) about
a 10 times smaller quality factor. In our measurement range, we do not
observe a significant 
dependence of the quality factor on doping concentrations.

\begin{figure}[htb]
\begin{center}
\includegraphics[width=10cm]{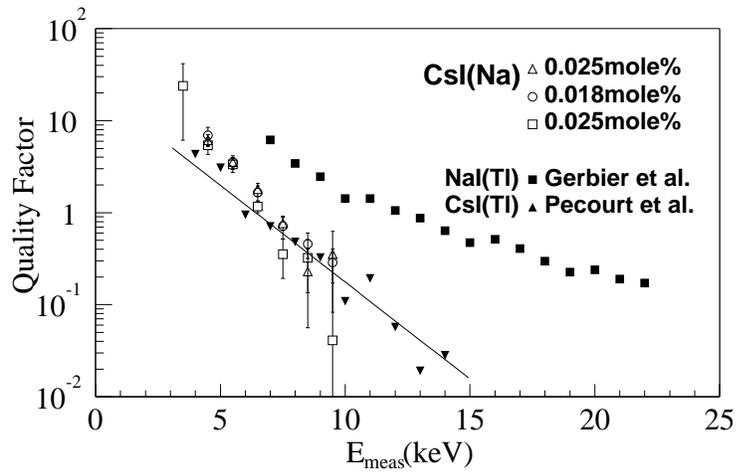}
\end{center}
\caption{Quality factors for various CsI(Na). The errors are only
statistical. }
\label{fig:NAquality}
\end{figure}

Figure \ref{fig:NAquality} shows quality factors for a CsI(Na) crystal.
The quality factor of CsI(Na) is
about 2 times larger than that of CsI(Tl), but still about 5-10 times
smaller than that of NaI(Tl). The quality factor for a CsI(Na) crystal
is about 2-3 times worse than CsI(Tl) crystal in the $E_{meas} <$
$\sim$7 keV region,  and is almost comparable above that region.

\section{Quenching factor}
\label{sec:QuenchingFactor}

\begin{figure}[htb]
\begin{center}
\includegraphics[width=10cm]{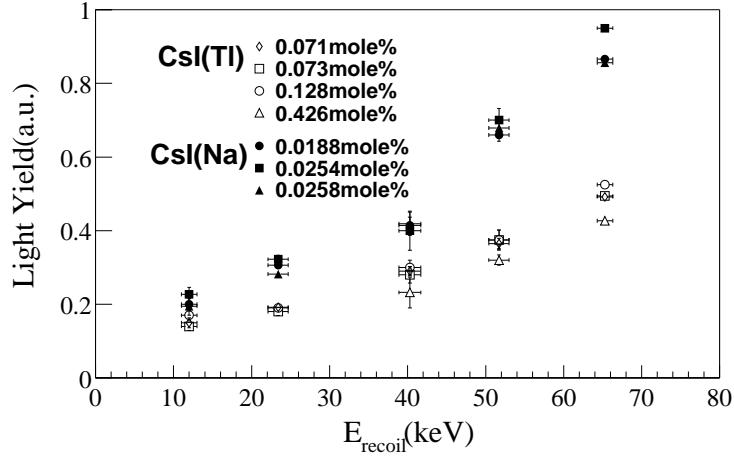}
\end{center}
\caption{Light yield vs recoil energy for CsI(Tl) and CsI(Na) with
different doping concentrations indicated in the figure.}
\label{fig:ErecoilvsLightYield}
\end{figure}

The light yield as a function of recoil energy($E_{recoil}$) for
various crystals is shown in Figure \ref{fig:ErecoilvsLightYield}. The
recoil energy is calculated by Eq. \ref{eq:Erecoil}. As expected,
CsI(Na) shows about a 30\% larger light yield than CsI(Tl) for the same
recoil energy. However, the light yield of the crystals with different
doping concentrations are consistent with each other within errors,
except for CsI(Tl-0.426mole\%), whose light transmission was not
good. It seems that the light yield is already saturated as indicated
by the doping concentration in our measurement range.

The energy calibration to obtain $E_{meas}$ uses a 5.9 keV X-ray
from $^{55}$Fe for the CsI(Tl) crystal. Non-linearity of the light
yield of the CsI(Tl) crystal at low energy\cite{ref:NonLinearity} is
considered as a systematic error of $E_{meas}$. For the CsI(Na) crystal,
the energy calibration uses a 59.5 keV X-ray from $^{241}$Am
since CsI(Na) is not sensitive to low energy photons, due perhaps to
some unknown surface effect. The response of CsI(Na) is assumed to be
linear within our measured energy range because the low energy
response($E_{meas} <$ 10 keV), including the non-linearity for CsI(Na), has
not yet been studied.

\begin{figure}[htb]
\begin{center}
\includegraphics[width=10cm]{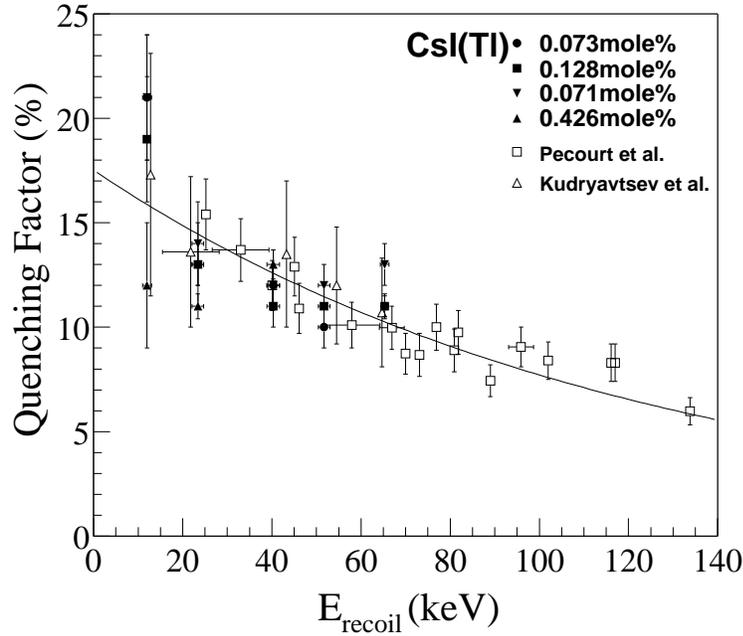}
\end{center}
\caption{Quenching factors for CsI(Tl) (present results:
closed markers). Previous measurements(open markers) are compared.} 
\label{fig:quenching}
\end{figure}

Using $E_{meas}$ and $E_{recoil}$, the quenching factor is calculated
to be;
\[Q = {E_{meas} \over E_{recoil}}.\]
Figure \ref{fig:quenching} shows that our measured quenching factors
for CsI(Tl) are in good agreement with previous
measurements\cite{ref:Pecourt,ref:Kudryavtsev} within errors. The
quenching factors for a Na-doped CsI crystal, which were measured for
the first time, are shown in Figure \ref{fig:quenching}. For the
comparison, the quenching factors for CsI(Tl-0.071mole\%) are shown
together. The quenching factor of CsI(Na) is about 50\% smaller than
that of CsI(Tl).

\section{Conclusion}

The intensive studies on the response of CsI(Tl) and CsI(Na) crystals
were performed at low energies ($ E_{meas} <$ 10 keV), and the
low energy response for a CsI(Na) crystal is reported for the first
time.  The pulse shape discrimination(PSD) between $\gamma$ and
nuclear recoil events was measured and the quality factor was
estimated using PSD.  The PSD and quality factor for CsI(Tl) are
similar to previous measurements. The quality factor
for a CsI(Na) crystal is about 2-3 times worse than for a CsI(Tl) crystal
in $E_{meas} <$ $\sim$7 keV region,  and almost comparable above
that region.

The quenching factor for CsI(Tl) and CsI(Na) crystals of various doping
concentrations were also measured.  As expected, CsI(Na) shows larger
light output than CsI(Tl) even in the low energy region. However, the
quenching factor of CsI(Na) for nuclear recoil is about a factor of
two smaller than that of CsI(Tl).

\section*{Acknowledgments}
This work is supported by Creative Research Initiatives Program of
Korean Ministry of Science and Technology. We thank Dr. K. D. Kim and
other staffs of Korea Institute of Geoscience and Mineral
Resources(KIGAM) for their support to the experiment. Also, we 
thank E. Hungerford for many helpful comments.

\newpage

\begin{table}[htb]
\caption{List of CsI crystals tested}
\label{tab:crystal}
\begin{tabular}{c c c c c } \hline
Doping Material           & Concentration(mole\%)  & Growing method \\ \hline
Tl                        & 0.073$\pm$0.02        & Bridgman\\ %0.15
Tl                        & 0.128$\pm$0.01       & Bridgman\\% 0.08
Tl                        & 0.426$\pm$0.23        & Bridgman\\ %0.30
Tl                        & 0.071$\pm$0.01        &
Kyropoulos\\  \hline
Na                        & 0.0188            & Kyropoulos\\ %0.021
Na                        & 0.0254            & Kyropoulos\\ %0.024
Na                        & 0.0258            & Kyropoulos\\ \hline %0.021
\end{tabular}
\end{table}

\newpage

\begin{table}[htb]
\caption{Timing characteristics of CsI crystals for nuclear
recoil(Errors are only statistical).}
\label{tab:timing}
\begin{tabular}{c c c c c c} \hline
Doping(mole\%)              & $\tau_r$($\mu$s) & $\tau_f$($\mu$s)  &$\tau_s$($\mu$s) & R & mean time($\mu$s) \\ \hline
Tl(0.073$\pm$0.02)  & $0.021\pm0.001$      & $0.546\pm0.007$  & $2.193\pm0.054$ &$0.426\pm0.011$ &$0.981\pm0.011$\\
Tl(0.128$\pm$0.01)  & $0.017\pm0.001$      & $0.560\pm0.008$  & $2.240\pm0.067$ &$0.393\pm0.012$  &$0.969\pm0.012$\\
Tl(0.426$\pm$0.23)   & $0.014\pm0.002$      & $0.523\pm0.020$  & $1.653\pm0.082$ &$0.582\pm0.024$ &$0.924\pm0.025$\\
Tl(0.071$\pm$0.01)  & $0.017\pm0.001$      & $0.457\pm0.017$  & $1.341\pm0.048$ &$0.763\pm0.021$  &$0.849\pm0.020$\\ \hline 
Na(0.0188)     & $0.035\pm0.002$      & $0.432\pm0.006$  &$2.437\pm0.050$ & $0.521\pm0.008$       &$1.037\pm0.018$\\ 
Na(0.0254)          & $0.029\pm0.001$      & $0.409\pm0.005$  &$2.422\pm0.046$ & $0.463\pm0.007$  &$0.966\pm0.012$\\ 
Na(0.0258)       & $0.032\pm0.001$      & $0.433\pm0.003$  &$2.753\pm0.036$ & $0.474\pm0.005$     &$1.038\pm0.013$\\ \hline 
\end{tabular}
\end{table}

\newpage

\begin{figure}[htb]
\begin{center}
\includegraphics[width=10cm]{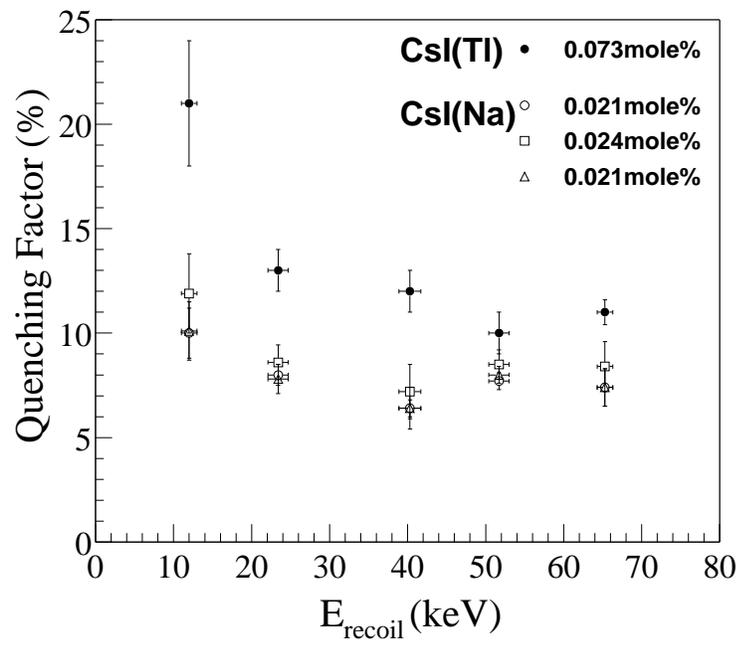}
\end{center}
\caption{Quenching factors for CsI(Na).}
\label{fig:NAquenching}
\end{figure}

\end{document}